\documentstyle [twocolumn,prb,aps] {revtex}
\begin{document}
\draft
\begin{title}
{Microscopic description of the surface dipole plasmon  
in large Na$_N$ clusters ($950 \lesssim N \lesssim 12050$)}
\end{title} 
\author{Constantine Yannouleas}
\address{
School of Physics, Georgia Institute of Technology,
Atlanta, Georgia 30332-0430 }
\date{ }

\maketitle
\begin{abstract}
Fully microscopic RPA/LDA calculations of the dipole plasmon for very large 
neutral and charged sodium clusters, Na$_N^{Z+}$, in the size range $950 
\lesssim N \lesssim 12050$ are presented for the first time. 60 different
sizes are considered altogether, which allows for an in-depth investigation 
of the asymptotic behavior of both the width and the position of the plasmon.
\end{abstract}

\pacs{PACS numbers: 36.40.Gk, 36.40.Vz}

\narrowtext

The surface dipole plasmon in large metal clusters is currently
attracting significant experimental \cite{kw,stein,kolw} and theoretical 
attention, \cite{bert,kura,nest} e.g., a time-resolved analysis \cite{kw} 
of second harmonic generation via femtosecond pump-probe studies was recently  
successful in establishing an upper limit of about 25 nm for the
validity of the $1/R$-law \cite{krei,yann1} (where $R$ is the cluster radius) 
associated with the width of the dipole plasmon in large Na$_N$ clusters.
On the theoretical side, this Landau-damping-type width, 
$\Gamma = A \hbar v_F/R$,
in large metal clusters has been derived repeatedly with various analytical
approaches \cite{krei,yann1} 
(which necessarily utilize simplified approximations for the dipole matrix
elements 
of the residual Coulomb force), but a microscopic justification of it is still
lacking. In general, detailed microscopic studies $\biglb($the most systematic 
among them using the matrix random phase approximation/local density
approximation (matrix-RPA/LDA) version 
\cite{yann2,yann3,yann4,yann5} of the linear response theory$\bigrb)$
have been resctricted to rather small clusters in the size range 
$8 \leq N \leq 338$. \cite{yann3,note3}

In the absence of systematic matrix-RPA/LDA calculations above $N=338$,
various other theoretical approaches have been most recently introduced
with the expressed purpose of providing a framework 
for microscopic studies of the
dipole plasmon in large metal clusters; they include the in-real-time 
version of the time-dependent local density approximation, \cite{bert,kura} 
the modified-with-separable-residual-forces RPA, \cite{nest} and the 
separation-of-collective-coordinates approach. \cite{kura} 
Due to the computational effort involved, however, these methods 
were in their turn restricted to a few 
(about 10) smaller sizes in the range \cite{kura} of $8 \lesssim N \lesssim 
400$ (or even in the range \cite{nest} $8 \lesssim N \lesssim 950$),
and as a result they
were unable to investigate the asymptotic behavior of the position and the
profile of the plasmon. In addition, through an identification of the
width with the variance of the RPA response, the authors of the recent
Rapid Communication 
\cite{kura} came via a semianalytical analysis to the conclusion that the 
plasmon width should exhibit an $1/\sqrt{R}$ asymptotic dependence, in sharp
contrast with the $1/R$ law.
   
In this paper, with the help of the new generation of powerful computers
available today, I present systematic matrix-RPA/LDA investigations of the 
dipole plasma excitation 
for neutral and charged sodium clusters in the size range $ 950 \lesssim N
\lesssim 12050$. In each case 30 different sizes were calculated, \cite{note} 
which allowed \cite{note9} for the determination of the asymptotic size 
dependence of both the position and the profile of the plasma resonance. 

The theoretical aspects of the matrix-RPA/LDA approach have been presented
in Ref.\ \onlinecite{yann3}. Here I only discuss briefly how the matrix-RPA
equations $\biglb($see Eq.\ (5) of Ref.\ \onlinecite{yann3}$\bigrb)$ 
can be reduced to an equivalent problem
with half the dimensions of the original matrix (such a reduction of dimensions
is instrumental for the successful implementation of the matrix-RPA/LDA
algorithm in the case of the very large metal clusters considered here). For 
such a reduction, it is sufficient to add and subtract the top and bottom
rows of the original matrix equation. 
One then finds the equivalent equation, \cite{note1}
\begin{equation}
D(2A^\lambda - D) Z^\nu ={\cal E}_\nu^2 Z^\nu~,
\label{modrpa}
\end{equation}
where the matrix $A^\lambda$ is given by Eq.\ (14) of Ref.\ \onlinecite{yann3},
$\lambda$ in general is the multipolarity of the excitation ($\lambda=1$ for
the case of the dipole),
${\cal E}_\nu$ are the RPA eigenvalues, and $D$ is a diagonal matrix depending
exclusively on the unperturbed single particle-hole energies, i.e.,
$D_{ph,p^\prime h^\prime}=(\epsilon_p - \epsilon_h) \delta_{pp^\prime}
\delta_{hh^\prime}$. The usual forward- and backward-going RPA amplitudes 
$W^\nu_{+} \equiv X^\nu$ and $W^\nu_{-} \equiv Y^\nu$, 
which are needed in order to calculate the oscillator strengths 
$f_\nu$ (see section II.C of Ref.\ \onlinecite{yann3}), 
are given through the eigenvectors $Z^\nu$ as follows:
\begin{equation}
W^\nu_{\pm} = {\cal C} \left( \frac{{\cal E}_\nu}{D} \pm \openone \right) 
Z^\nu~,
\end{equation}
where the proportionality constant ${\cal C}$ is determined via the 
RPA-eigenvector normalization condition.

For each cluster Na$_N^{Z+}$, the microscopic RPA calculation yields a
discrete set of oscillator strengths $f_\nu$ (normalized to unity) 
associated with the RPA eigenvalues ${\cal E}_\nu$
(see inset of Fig.\ 1). By folding the oscillator
strengths with Breit-Wigner shapes normalized to unity, one can calculate
a smooth photoabsorption cross section, $\sigma$, per valence electron as
follows:
\begin{equation}
\sigma(E) = 1.0975 \; ({\text{eV \AA}}^2) \sum_\nu 
f_\nu P_{\text{BW}}(E; \; {\cal E}_\nu,\gamma_\nu)~,
\label{sig}
\end{equation}
where $\gamma_\nu$ denotes the intrinsic widths of the auxiliary
Breit-Wigner profiles, $P_{\text{BW}}$.

Fig.\ 1 displays the calculated $\sigma$'s for Na$_{952}$
and Na$_{12068}$, namely for the smallest and the largest size considered in 
the present paper (intrinsic widths of $\gamma_\nu = \gamma=0.16$ and 0.076 eV 
were used for Na$_{952}$ and Na$_{12068}$, respectively). 
It is apparent that the FWHM (full width at half maximum) is smaller for 
Na$_{12068}$ ($\sim $ 0.147 eV) compared to the case of Na$_{952}$
($\sim$ 0.319 eV) and that 
the maximum of the photoabsorption cross section for $N=12068$ is blueshifted 
with respect to the case of $N=952$. 

In addition to the photoabsorption
profiles, Fig.\ 1 also displays the variance $\Sigma$ (horizontal solid 
bar) associated with the RPA dipole response in the case of Na$_{12068}$
$\biglb($in the RPA 
the variance is calculated \cite{brac} via the relation 
$\Sigma^2=(m_2/m_0)-(m_1/m_0)^2$, where the moments $m_k$ are given by 
$m_k=\sum_\nu {\cal E}_\nu^{k-1}f_\nu$$\bigrb)$. 
Even with the overestimation of the
FWHM due to the folding procedure (see below), it is seen that the variance
is substantially larger than the corresponding FWHM in the case of 
Na$_{12068}$; thus the variance should not be used as a 
substitute for the actual width of the dipole plasmon, as was
done \cite{note10} in Ref.\ \onlinecite{kura}. 
Furthermore, to correct for the overestimation due to the folding, one
needs to subtract the intrinsic width $\gamma_\nu$ from the original FWHM
(see Fig.\ 10 of Ref.\ \onlinecite{yann3} and accompanying discussion). 
With this correction, the plasmon
widths are $\sim $ 0.16 eV and $\sim$ 0.07 eV for Na$_{952}$ and Na$_{12068}$,
respectively; these values are even smaller than the uncorrected FWHM's
compared to the corresponding variances. 

The method of folding described above has been used in most of the earlier 
publications in order to extract the FWHM of the photoabsorption profiles.
This method, however, becomes cumbersome in the present study where the 
size changes substantially going from the smallest to the largest cluster.
Indeed, smaller $\gamma_\nu$'s must be used the larger the 
cluster size, and such a procedure is difficult to be determined uniquely.
Instead of the folding method, a more uniform procedure
is needed. Using the fact that for large cluster sizes it can be analytically 
shown \cite{ydg} that the RPA response itself tends asymptotically to a well 
concentrated \cite{note4} distribution of oscillator strengths $f_\nu$ 
exhibiting a Breit-Wigner profile, i.e.,
\begin{equation}
f_\nu \propto \Gamma / [({\cal E}_\nu - \overline{\cal E})^2 + (\Gamma/2)^2]~, 
\label{eq4}
\end{equation}
this paper has opted for a method based
directly on the distribution of the $f_\nu$'s; namely,
the Landau-damping-type width $\Gamma$ is determined by the minimum 
interval around the RPA energy centroid $\overline{\cal E}$ which contains 
50\% of the total oscillator strength, i.e., $\Gamma$ is such that 
$\sum_i f_i=0.5$ with ${\cal E}_i \in (\overline{\cal E}-\Gamma/2,
\overline{\cal E}+\Gamma/2)$. The value of 50\% used here is precisely the
percent of the total oscillator strength contained within the energy interval 
associated with the FWHM of the single Breit-Wigner profile [see Eq.\
(\ref{eq4})], and the RPA centroid is given by $\overline{\cal E}=m_1/m_0$.

Fig.\ 2(a) displays the calculated (according to the latter method)
RPA widths ($\Gamma$, open circles) and 
variances ($\Sigma$, open squares) for Na$_N$ clusters in the size range 
$ 952 \leq N \leq 12068$ as 
a function of the inverse radius ($1/R$) of the jellium background 
($R=r_sN^{1/3}$, and the Wigner-Seitz radius for sodium was taken as 
$r_s=4$ a.u.). Fig.\ 2(b) displays the same quantities as a function of
$1/\sqrt{R}$. Before proceeding further, it needs to be noticed that the
corrected FWHM's for Na$_{952}$ and Na$_{12068}$ (namely the values 0.16 
and 0.07 eV) according to the folding method
are in good agreement with the widths $\Gamma$ derived from the 50\% method.

A first observation is that the variances are systematically
larger than the widths (see also Fig.\ 1). Second, it can be seen that the
widths are not proportional to $1/\sqrt{R}$, since the
associated straight line does not pass through the origin of the axes 
[Fig.\ 2(b)]. On the contrary, in the case of Fig.\ 2(a), the 
straight dashed line
that passes on the average through the points denoting the widths
definitely passes through the origin of axes. This provides an unequivocal
proof that $\Gamma \propto 1/R$. Notice that for the smaller sizes 
the precise values of the widths exhibit noticeable scattering around the 
average line specified by the $1/R$ law; however, such scattering becomes 
progressively smaller the larger the cluster sizes. The solid line in Fig.\ 
2(a) denotes the $\Gamma=A \hbar v_F/R$ law according to the analytical 
result of Ref.\ \onlinecite{yann1}, namely when \cite{note5} 
$A=A_{\text{ana}}= 0.46 $. For the numerical RPA calculation,
one finds $A_{\text{num}}=0.58$, which is close to $A_{\text{ana}}$.
The difference between the two slopes is due to the infinite-well
approximation (and ensuing neglect of electronic spillout) which was invoked
during the analytical derivation. 
To further test the $1/R$ dependence, I have also plotted in Fig.\ 2(a) the 
{\it uncorrected} FWHM's of the photoabsorption profiles for Na$_{952}$ and
Na$_{12068}$ (solid dots). Again, and in sharp contrast with the variances, 
even these uncorrected FWHM's follow clearly an $1/R$ dependence.

Finally, it has been derived through 
semiclassical arguments \cite{kura,brac} that the size dependence of the
variance should be propotional to $1/\sqrt{R}$. Fig.\ 2(b) does not indicate
such a relation; much larger sizes are needed before the microscopic results
for $\Sigma$ do converge to the expected $1/\sqrt{R}$ relation.

Fig.\ 3 portrays the size evolution of the quantities associated with the
position of the RPA surface plasmon, namely, it displays the 
quantities ${\cal E}_1=(m_1/m_{-1})^{1/2}$, ${\cal E}_3=(m_3/m_1)^{1/2}$, and
the centroid $\overline{\cal E}$ as a function of the inverse cluster radius.
Fig. 3(a) corresponds to the case of neutral Na$_N$ clusters, while Fig.\ 3(b)
displays the case of multiply cationic Na$_N^{10+}$ clusters. 
In all instances, the 
calculated RPA values honor the theoretically expected \cite{note6} inequality
${\cal E}_1 \leq \overline{\cal E} \leq {\cal E}_3$, but the centroid lies
much closer to ${\cal E}_1$ than to ${\cal E}_3$. For the neutral clusters, 
it is seen that the calculated points for ${\cal E}_1$ and
$\overline{\cal E}$ lie closely on two straight lines, which cross the 
energy axis ($R=\infty$) at about 3.403 eV, in very good agreement with the
expected classical Mie limit for sodium clusters, i.e, $\hbar 
\omega_{\text{Mie}} = \hbar \omega_{\text{bulk}}/\sqrt{3} = 3.4$ eV (for 
$r_s=4$ a.u.). From the relation, $\overline{\cal E} =3.403 \; \text{eV} -
\overline{\eta}/R$, one can determine $\overline{\eta}=4.68$ (eV a.u.). 
The calculated points for ${\cal E}_3$ 
do not fall on a straight line crossing the
energy axis at about 3.4 eV, although such a behavior has been
predicted from semiclassical sum-rules arguments; \cite{brac,note7} apparently
much larger sizes are needed before the size dependence of ${\cal E}_3$
converges to this expected asymptotic behavior.

Comparing the RPA results in Fig.\ 3(b) with corresponding
results in Fig.\ 3(a), one sees that the position of the dipole plasmon in 
the case of charged clusters is blueshifted with respect to that of the neutral
clusters (all three quantities ${\cal E}_1$, $\overline{\cal E}$, and
${\cal E}_3$ exhibit this trend). The extent of this blueshift depends on the
ratio $Z/N$, and naturally for the larger 
charged clusters the calculated positions
tend to converge to the corresponding results for the neutral clusters; for 
$Z=10$, this convergence is definitely recognizable for the largest sizes 
studied here. 

Finally, Fig.\ 4 displays the size evolution of the RPA widths $\Gamma$ and 
variances $\Sigma$ of the charged Na$_N^{10+}$ clusters as a
function of $1/R$. As was the case with the position,
the charge state influences the values of $\Gamma$ and $\Sigma$; in particular
a reduction in magnitude of these quantities can be seen compared to the
case of neutral clusters. This reduction depends again on the ratio
$Z/N$, and it is more pronounced the smaller the cluster size. \cite{note8}

In conclusion, fully microscopic RPA/LDA calculations for neutral Na$_N$ and 
charged Na$_N^{10+}$ ($ 950 \lesssim N \lesssim 12050 $) clusters 
were presented. In each case, some 30 different
sizes were considered, which allowed for an in-depth investigation of the
asymptotic behavior of both the position and the width of the dipole plasmon.
In particular, it was found that asymptotically the RPA width 
becomes proportional to the inverse cluster radius, and that for neutral
clusters this trend is already well developed within the size range considered
here.  

This research is supported by the USDOE and was performed at the
GIT Center for Computational Materials Science.

\begin{figure}
\caption{
RPA photoabsortion cross sections for Na$_{952}$ (dashed line) and Na$_{12068}$
(solid line). The solid horizontal bar denotes the RPA variance $\Sigma$
in the case of Na$_{12068}$. 
Inset: the oscillator-strength distribution for Na$_{12068}$.
}
\end{figure}

\begin{figure}
\caption{
RPA widths $\Gamma$ (open circles) and variances $\Sigma$ 
(open squares) of the dipole plasmon for
neutral Na$_N$ clusters ($952 \protect\lesssim N \protect\lesssim 12068$),
as a function of
(a) the inverse cluster radius, $1/R$ and (b) the inverse of the square root 
of the radius, $1/\sqrt{R}$. The solid line in (a) corresponds to the 
analytical result for $\Gamma$ according to Ref.\ \protect\onlinecite{yann1}.
The solid dots in (a) denote the {\it uncorrected\/} FWHM of the 
photoabsorption profiles (see Fig.\ 1) for Na$_{952}$ and Na$_{12068}$
generated with the folding method.
}
\end{figure}

\begin{figure}
\caption{
The RPA positional quantities ${\cal E}_1$ (open squares), $\overline{\cal E}$
(centroid, solid circles), and ${\cal E}_3$ (open triangles) 
of the dipole plasmon, as a function
of $1/R$ for (a) neutral Na$_N$ and (b) charged Na$_N^{10+}$ clusters. The
straight line in (b) is a duplicate of the lowest line in (a) and was drawn 
as a guide to the eye.
}
\end{figure}

\begin{figure}
\caption{
RPA widths $\Gamma$ (open circles) and variances $\Sigma$ (open squares) of 
the dipole plasmon for charged Na$_N^{10+}$ clusters 
($962 \protect\lesssim N \protect\lesssim 12078$), as a function of the 
inverse cluster radius, $1/R$. The straight lines are duplicates of dashed
lines in Fig.\ 2(a) and were drawn as guides to the eye.
}
\end{figure}


\begin{references}
\bibitem{note}
In the case of neutral (charged) clusters, Na$_{952}$ (Na$_{962}^{10+}$) 
was the smallest and Na$_{12068}$ (Na$_{12078}^{10+}$) the largest cluster 
studied. In all instances, a spherical jellium background was assumed for 
simplicity. The energy-weighted sum rule was fulfilled with a better than
0.1\% accuracy.
\bibitem{note1}
In the derivation of Eq.\ (\protect\ref{modrpa}), the relation
$A^\lambda - D = (-)^\lambda B^\lambda$ (valid in the case of the Coulomb
residual force) was used to eliminate the RPA submatrix $B^\lambda$.
\bibitem{note4}
For small clusters, e.g., Na$_8$, Na$_{20}$, and Na$_{40}$,
the fragmentation of the oscillator strength can be irregular and can lead
to the appearance of multipeak photoabsorption profiles (see Refs.\ 
\onlinecite{yann2,yann3}).
\bibitem{note7}
No such closed semiclassical asymptotic behavior has been derived for 
${\cal E}_1$ and $\overline{\cal E}$.
\bibitem{note8}
A similar dependence of the plasmon width (and of the position as discussed 
earlier) on the charge state was also found in the 
case of small clusters with $N-Z \leq 40$ 
(see Refs.\ \onlinecite{yann2,yann4,yann5}).
\end{references}
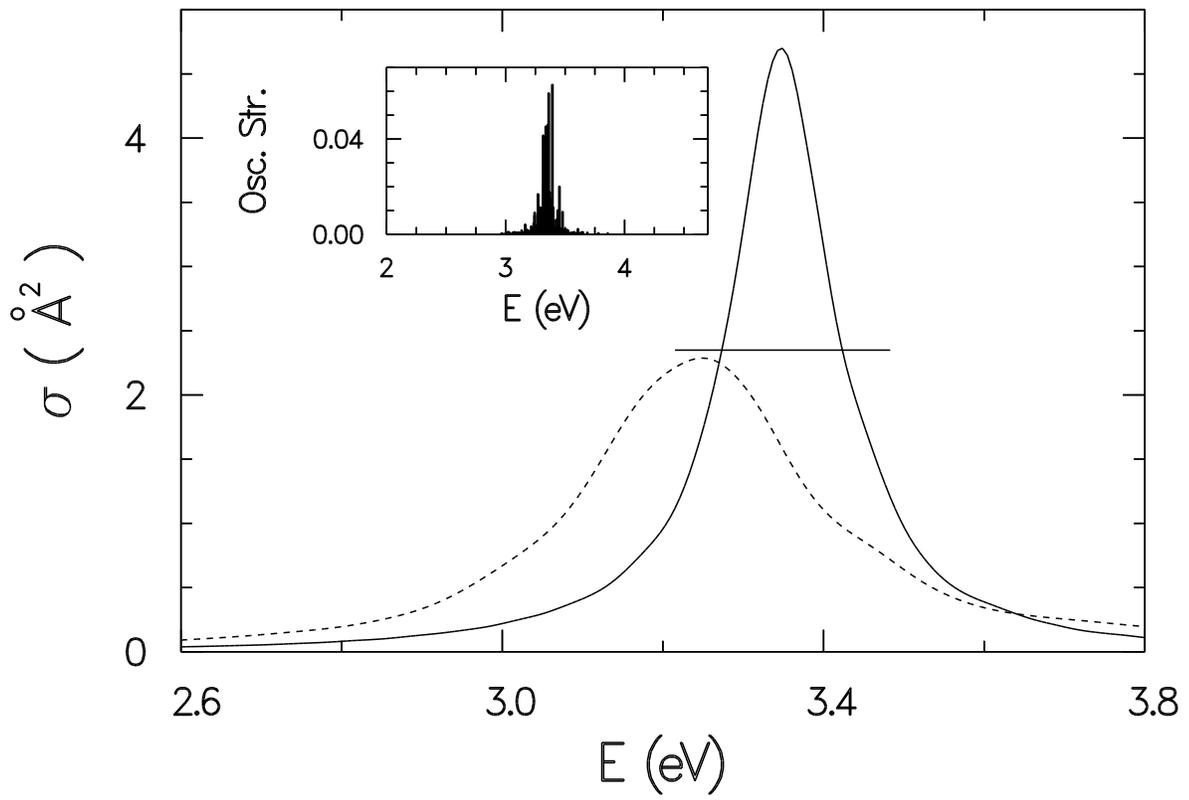
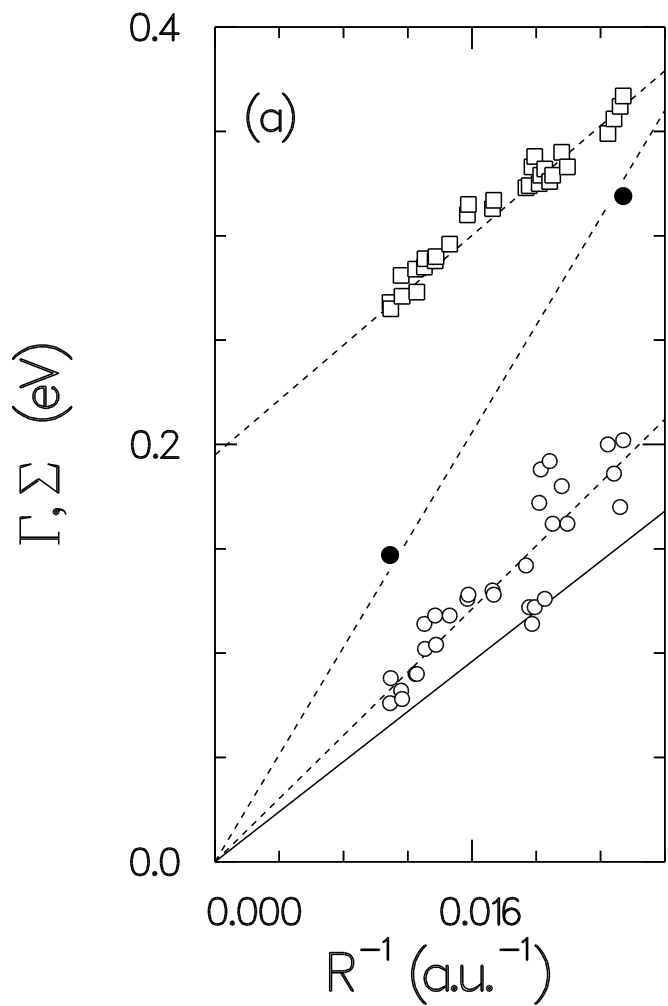
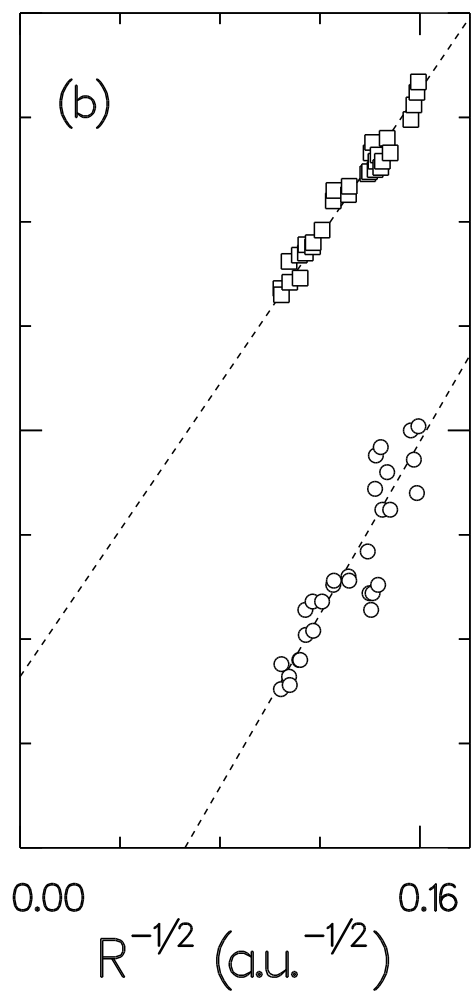
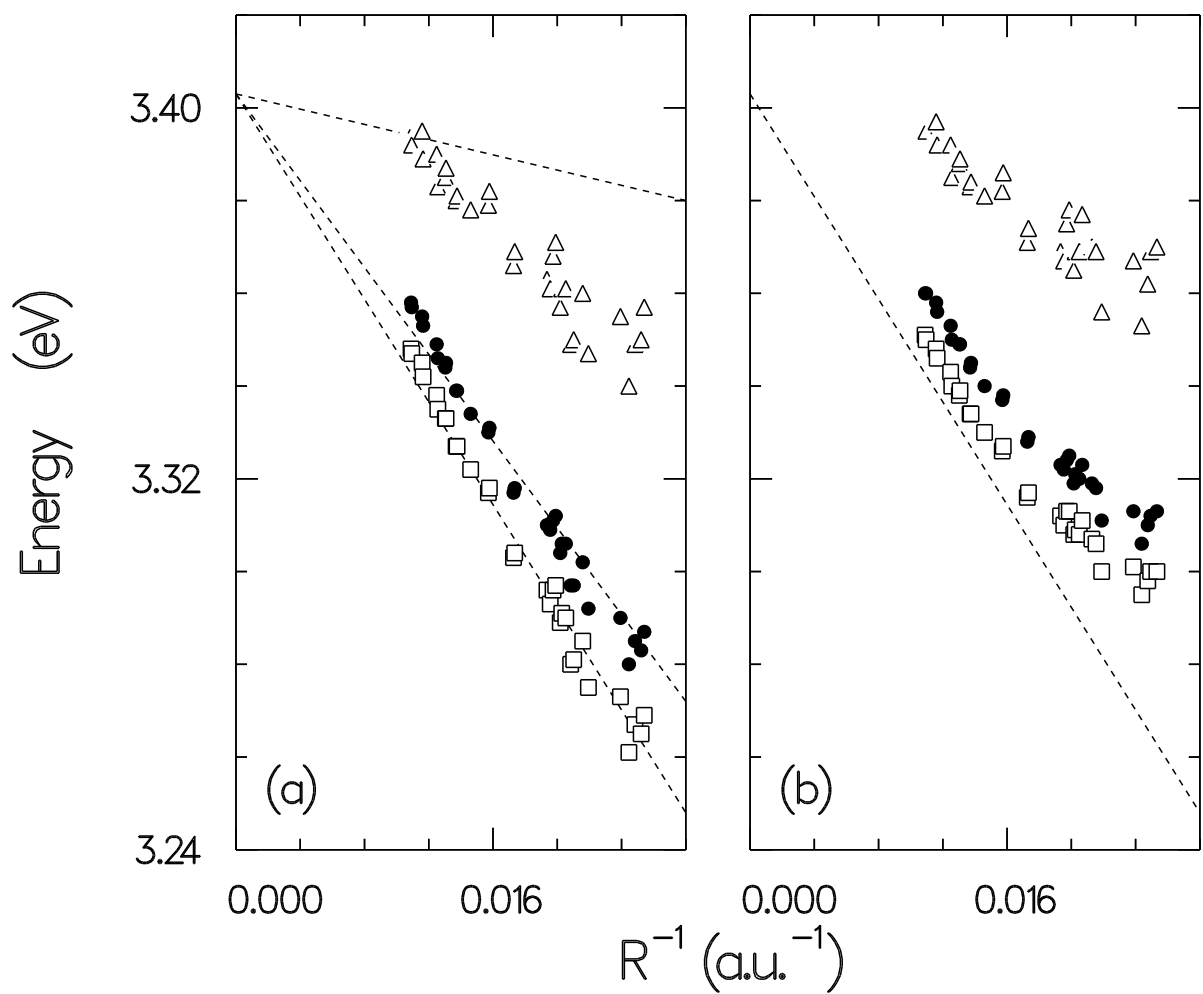
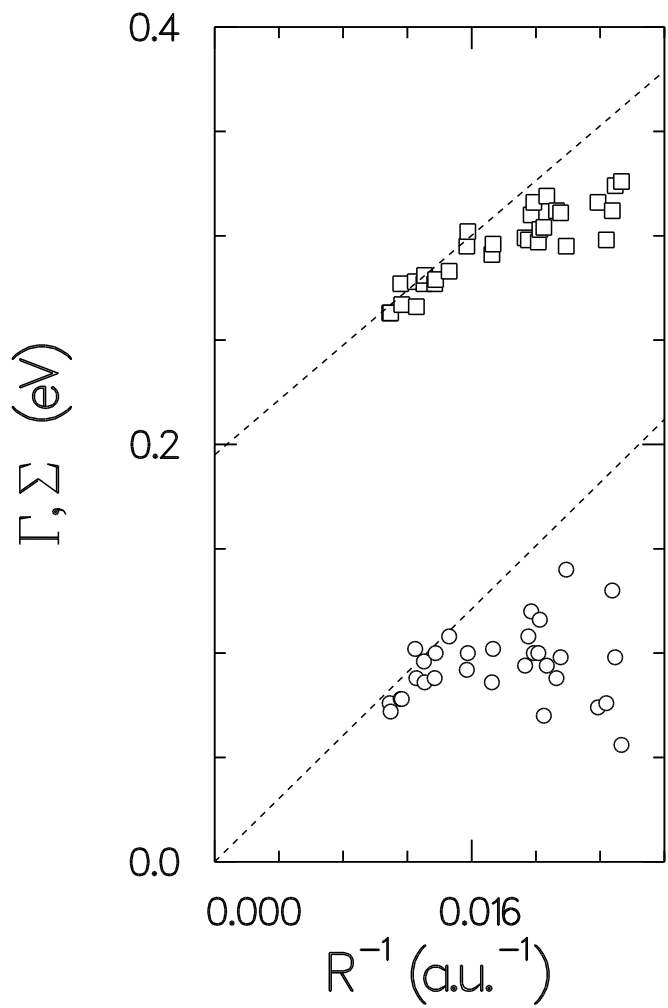
\end{document}